\documentclass[aps, twocolumn, prl, showpacs, superscriptaddress]{revtex4}
\usepackage{CJK}
\usepackage{graphicx}

\usepackage{dcolumn}
\usepackage{bm}
\usepackage{amsmath}

\begin{document}

\title{Interaction induced mergence of Dirac points in Non-Abelian optical lattices}
\author{Li Wang and Libin Fu$^*$ }

\affiliation{Institute of Applied Physics and Computational Mathematics, Beijing 100088, China}

\date{\today }

\email[E-mail: ]{ lbfu@iapcm.ac.cn }

\begin{abstract}
We study the properties of an ultracold Fermi gas loaded in a square
optical lattice and subjected to an external and classical non-Abelian
gauge field. We calculate the energy spectrum of the system and show that the
Dirac points in the energy spectrum will remain quite stable under on-site
interaction of certain strength. Once the on-site interaction grows stronger
than a critical value, the Dirac points will no longer be stable and merge into a single
hybrid point. This mergence implies a quantum phase transition from a semimetallic phase
to a band insulator. The on-site interaction between ultracold fermions could be conveniently
controlled by Feshbach resonances in current experiments. We proposed that
this remarkable interaction induced mergence of Dirac points may
be observed in the ultracold fermi gas experiments.

\end{abstract}

\pacs{67.85.Lm,37.10.Jk,71.10.Fd} \maketitle

One of the most interesting properties of graphene \cite{Novoselov},
a single layer of carbon atoms packed in a honeycomb lattice, lie in
the fact that the low energy excitations obey a linear dispersion
relation \cite{Wallace} around the so-called Dirac points, and thus
can be used as a testbed for the relativistic quantum
electrodynamics. Consequently, it is now possible to observe many
remarkable phenomena in table-top experiments, such as Klein
tunneling \cite{Katsnelson, Stander} and the relativistic extension
of Landau levels \cite{Rabi, McClure, Li}, which usually only occur
in high-energy physics\cite{Neto}. This advantage of graphene has
stimulated a great interest in the investigation of Dirac points
\cite{Neto} in many other systems.  In particular, ultracold atoms in
optical lattices provide a versatile playground where the properties
of condensed matter systems can be simulated \cite{Lewenstein,
Bloch} in a highly controllable manner, such as the superfluid-Mott
insulator transitions of Hubbard models \cite{Greiner, Jordens,
Schneider}. A quantum-optical analogue of graphene can be achieved
by loading ultracold fermionic atoms such as $^{40}$K or $^{6}$Li
\cite{Jordens, Schneider} into a hexagonal optical lattice
\cite{Zhu}. The effects of Dirac points were discussed in the context
 of ultacold atoms in honeycomb lattice \cite{Zhu} and $T_3$ (rhombic)
 lattices \cite{Bercioux}. Moreover, much more intriguing phenomena
 arise when these systems are subjected to artificial non-Abelian
 gauge fields \cite{Lin, Osterloh, Ruseckas}, such as the non-Abelian
 Aharonov-Bohm effect\cite{Osterloh}, non-Abelian atom optics
 \cite{Juzeliunas}, quasirelativistic effects \cite{Juzeliunas1}, or
 exotic topological phase transitions \cite{Bermudez}.

 Here we would like to emphasize a fact that the non-Abelian artificial gauge field
 also provide an interesting setup where Dirac points emerge in a square
 optical lattice \cite{Goldman}, which is originally limited to staggered
 fields \cite{Lim, Hou}. In this article, we consider a similar system in
 which two-component (two-color) ultracold fermionic atoms are trapped in a
 square optical lattice. And dramatic difference between the two systems
 comes from the repulsive on-site interaction introduced into the model by us. Works
 \cite{Bermudez,Goldman} mentioned above mainly dwelled on free Fermi gas on 2D
  optical lattice. We study the effects of repulsive on-site interaction on the
  energy spectrum of the 2D fermi gas loaded into a square optical lattice and
  subjected to a non-Abelian artificial gauge field. Implementing a self-consistent
  mean-field theory, we show that the Dirac points in the energy spectrum remain quite stable
  under repulsive on-site interaction of certain strength. When the on-site interaction grows
  stronger than a critical value, the Dirac points will no longer be stable and two Dirac points merge into a single one. This merging indicates a quantum phase transition between
  a semimetallic phase and a band insulator \cite{Zhu, Hasegawa, Dietl, Goerbig, Pereira, Wunsch, Volovik, Montambaux}. And one thing worthy to be mentioned here
  is that once the two Dirac points merge, the final dispersion relation becomes quite
  exotic---it is linear in one direction but parabolic in the other orthogonal direction.
  Very recently, a well-designed experiment has been carried out by Leticia \emph{et al}. \cite{Leticia}, in which the creating, moving and  merging of Dirac points has been realized with a Fermi gas loaded into a tunable honeycomb lattice. While the creating, moving and merging of Dirac points in the experiment \cite{Leticia} is generated by designing complex lattice geometries, the mergence of the Dirac points in our model is induced by strong repulsive on-site interaction. Since pairwise interactions could be conveniently controlled by means of Feshbach resonances \cite{Timmermans}, we propose that the interesting mergence of Dirac points in our model may be experimentlly observed and characterized in non-Abelian optical lattices.
  The non-Abelian optical lattice could be prepared by generalizing the recent experiment \cite{Lin}, as proposed in \cite{Osterloh,Ruseckas}.

\begin{figure}[tbp]
\setlength{\unitlength}{2mm}
\begin{center}
\begin{picture}(40,20)
\put(0,0){\scalebox{0.5}[0.6]{\includegraphics[width=8.5cm]{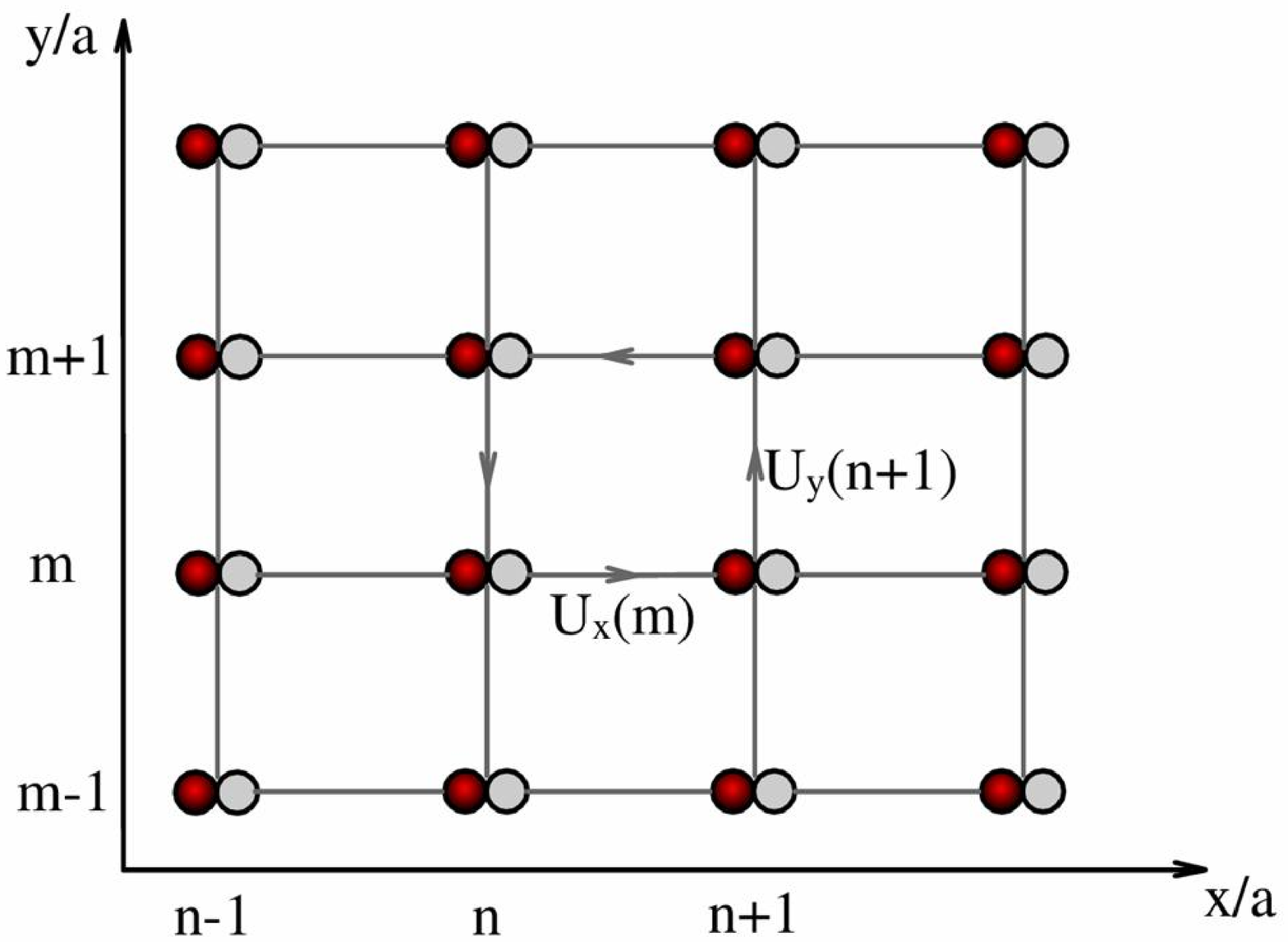}}}
\put(18,0){\scalebox{0.5}[0.5]{\includegraphics[width=8.5cm]{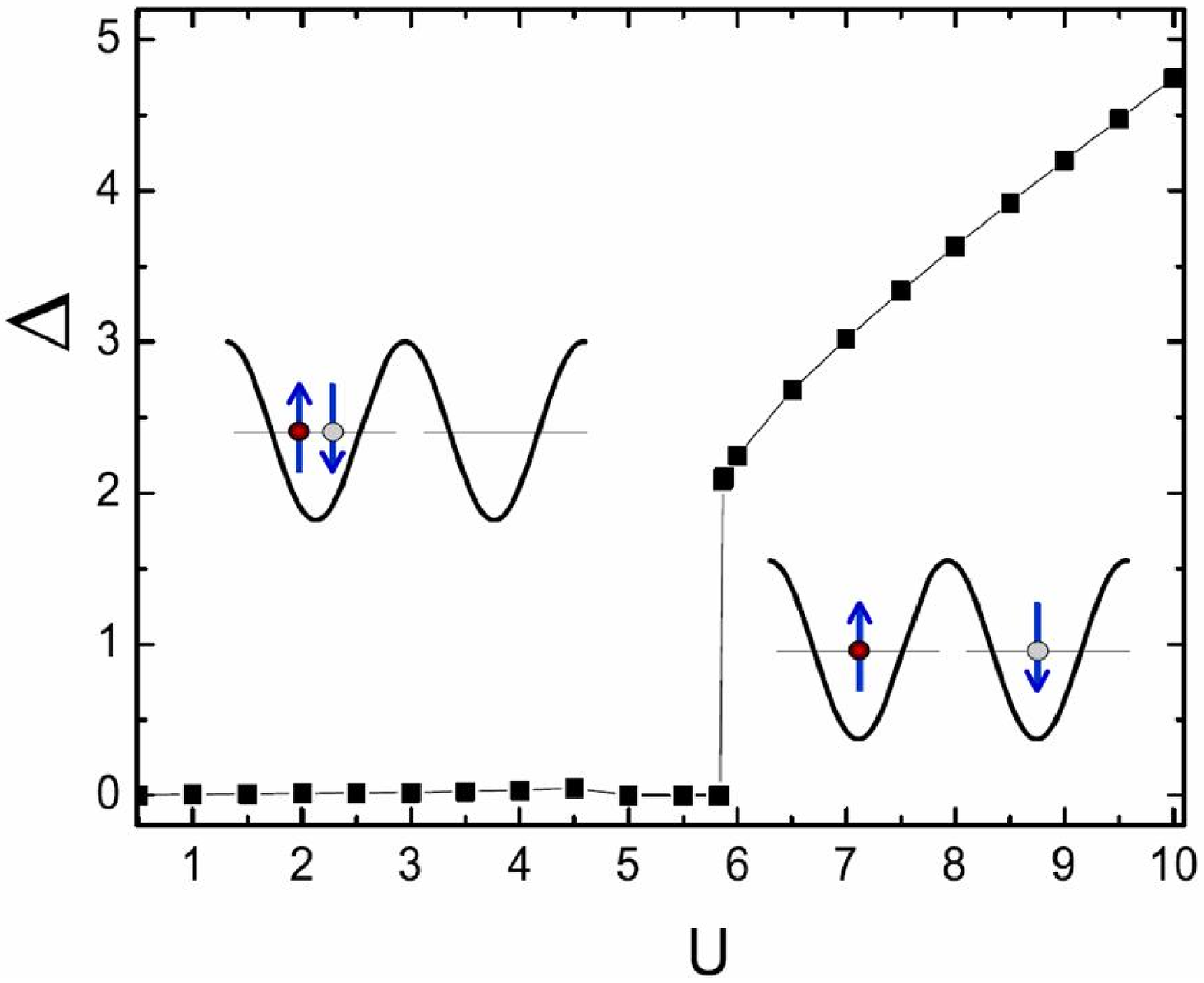}}}
%\put(24,8){\scalebox{0.2}[0.2]{\includegraphics[width=6cm]{state1.eps}}}
%\put(33,4){\scalebox{0.2}[0.2]{\includegraphics[width=6cm]{state2.eps}}}
\put(3,17.5){(a)}
\put(19.5,17.5){(b)}
\end{picture}
\end{center}
\caption{(color online) (a) Two-component (two-color) ultracold fermionic atoms trapped in a square optical lattice and subjected to an non-Abelian artificial gauge field. The circle filled with red(gray) color denotes internal atom state $ \left| \uparrow \right > $  $ (\left| \downarrow \right >)$. $U_x(m)$ and $U_y(n+1)$ are unitary operators induced by the external artificial gauge field. (b) The self-consisted mean-field order parameter $\Delta$ vs on-site interaction $U$. The system size
is $24 \times 24$ and we set the convergence criterion as $10^{-4}. $ \label{lattice}}
\end{figure}

We consider a two-component (two-color) Fermi gas trapped on a square optical lattice and subjected to an artificial non-Abelian gauge potential. Fermionic atoms in the system are interacting with repulsive on-site interaction. It is well known that this pairwise interactions can be freely tuned by means of Feshbach resonances \cite{Timmermans} in nowadays ultacold atoms experiments. The Hamiltonian of the system reads,

\begin{equation}
H=-t\sum_{<\mathbf{r}\mathbf{r'}>}\sum_{\tau,\tau'}(c_{\mathbf{r}\tau}^{\dagger}e^{-i\int_{\mathbf{r'}}^{\mathbf{r}}\mathbf{A} \cdot d\mathbf{l}}c_{\mathbf{r'}\tau'}+H.c.)+V\sum_{\mathbf{r}}n_{\mathbf{r} \uparrow}n_{\mathbf{r}\downarrow} , \label{Ha}
\end{equation}
where $t$ is the hopping amplitude, $c_{\mathbf{r},\sigma}$($c_{\mathbf{r},\sigma}^{\dagger}$) is the fermionic annihilation (creation) operator at site $\mathbf{r}$ of the square optical lattice, $\tau=\uparrow,\downarrow$ can be regarded as pseudospin index, $\left<\mathbf{r}\mathbf{r'}\right>$ denotes that the sum is over nearest neighbors and $V$ is the strength of the on-site interaction between fermionic atoms. The coordinate of a fermion is given by $\mathbf{r}=(ma,na)$, where $m,n$ are integers and $a$ is the lattice constant of the square optical lattice. Here we set $\hbar=e=1$. The external gauge potential has the following form, $\mathbf{A}=\frac{B_0}{2}(-y,x)+a(B_{\alpha}\sigma_y,B_{\beta}\sigma_x)$, in which $B_0,B_\alpha,B_\beta$ are experimentally controllable parameters and $\sigma_{x,y}$ are the Pauli matrices.
This intriguing artificial gauge field can be realized following the proposals \cite{Osterloh, Ruseckas, Goldmanpra}, along the lines of the recent experiment \cite{Lin}. After some algebra, the original Hamiltonian (\ref{Ha}) becomes
\begin{equation}
H=-t\sum_{<\mathbf{r}\mathbf{r'}>}\sum_{\tau,\tau'}(c_{\mathbf{r}\tau}^{\dagger}[U_{\mathbf{r}\mathbf{r'}}]_{\tau\tau'}c_{\mathbf{r'}\tau'}+H.c.)+V\sum_{\mathbf{r}}n_{\mathbf{r} \uparrow}n_{\mathbf{r}\downarrow} , \label{Hb}
\end{equation}
where $U_{\mathbf{r}\mathbf{r'}}$ is the matrix form of a nontrivial unitary operator accompanying the hopping between nearest neighbor $\mathbf{r}$ and $\mathbf{r'}$. If the hopping is along $x$ axis, $U_{\mathbf{r}\mathbf{r'}}=U_x(m)=e^{-i\pi \Phi m} e^{i\Phi_\alpha \sigma_y}$. If the hopping is along $y$ axis, $U_{\mathbf{r}\mathbf{r'}}=U_y(n)=e^{-i\pi \Phi n} e^{i\Phi_\beta \sigma_x}$. $\Phi=B_0a^2$ is the Abelian magnetic flux, and $\Phi_{\alpha,\beta}=B_{\alpha,\beta}a^2$ is the non-Abelian flux. Fermions hopping around an elementary square indicate a unitary transformation \cite{Goldman} $U=U_x(m)U_y(n+1)U_x^\dagger(m+1)U_y^\dagger(n)$. The boundary between Abelian regime and non-Abelian regime is well defined by the gauge-invariant Wilson loop\cite{Goldman,Goldmanpra} $W=trU$. Here we constrain ourselves to the non-Abelian regime, where the Wilson loop $|W|<2$ and we set the Abelian flux $\Phi=0$.

In the noninteracting limit of Hamiltonian (\ref{Ha}), i.e. the case in which fermions hop freely between neighbor sites without any interaction, Hamiltonian of the system is a beautiful quadratic form and can be analytically solved by Bogoliubov transformations. Energy spectrum of this case has been beautifully analyzed in literature \cite{Goldman}, where the fermion gas becomes a collections of noninteracting quasiparticles and the spectrum develops four independent Dirac points at $K_D \in \{(0,0),(\frac{\pi}{a},0),(0,\frac{\pi}{a}),(\frac{\pi}{a},\frac{\pi}{a})\}$ in the vicinity of marginally Abelian regime ($\Phi_\alpha, \Phi_\beta \approx \pi/2$). However, once the on-site interaction is taking into account in Hamiltonian (\ref{Ha}), it's not a quadratic form any more and therefore can't be solved by Bogouliubov transformation directly. This is right the case we consider in this article. We study the repulsively interacting fermions on a square optical lattice subjected to a non-Abelian gauge field by the means of a self-consistent mean field theory. Our startpoint is Eq. (\ref{Hb}).

We mainly consider the repulsive interaction regime in this paper. As the on-site interaction grows stronger and stronger, fermionic atoms with different colors on the square lattice tend to repel each other and avoid staying on the same site. Once the interaction strength is over a critical point, the square optical lattice at half-filling will enter a phase in which each site of the lattice is occupied by single atom (see Fig. \ref{lattice}b). Therefore, we define $\Delta_{\mathbf{r}}=V \left<c_{\mathbf{r}\uparrow}^\dagger c_{\mathbf{r}\downarrow}\right>$ as our order parameter. Under this mean-field approximation, the Hamiltonian (\ref{Hb}) can be written as a quadratic form,

\begin{align}
H_{MF}=&-t\sum_{<\mathbf{rr'}>}\sum_{\tau\tau'}(c_{\mathbf{r}\tau}^\dagger[U_{\mathbf{rr'}}]_{\tau\tau'}c_{\mathbf{r'}\tau'}+H.c.)  \nonumber \\
&-\sum_{\mathbf{r}}(\Delta_{\mathbf{r}} c_{\mathbf{r}\downarrow}^\dagger c_{\mathbf{r}\uparrow}+H.c.)+ \nonumber \\
&+\frac{VN}{2}+\frac{1}{V}\sum_{\mathbf{r}}|\Delta_{\mathbf{r}}|^2 .    \label{Hc}
\end{align}
Through a canonical transformation, the above Hamiltonian can be diagonalized by solving the following BdG equation \cite{bdg}:
\begin{equation}
\sum_{\mathbf{r'}}\left(
                          \begin{array}{cc}
                            h_{\mathbf{rr',\uparrow}} & O_{\mathbf{rr'}} \\
                            O_{\mathbf{rr'}}^* & h_{\mathbf{rr',\downarrow}} \\
                          \end{array}
\right) \left(
          \begin{array}{c}
            u_{\mathbf{r'}}^n \\
            v_{\mathbf{r'}}^n \\
          \end{array}
        \right)
=E_n \left(
          \begin{array}{c}
            u_{\mathbf{r'}}^n \\
            v_{\mathbf{r'}}^n \\
          \end{array}
        \right)
\end{equation}
where $h_{\mathbf{rr',\tau}}=-t[U_{\mathbf{rr'}}]_{\tau\tau}$, $O_{\mathbf{rr'}}=-\Delta_{\mathbf{r}}\delta_{\mathbf{rr'}}-t[U_{\mathbf{rr'}}]_{\uparrow\downarrow}$ and ($u_{\mathbf{r'}}^n, v_{\mathbf{r'}}^n $) are the eigenvectors correspoinding to the eigenenergy $E_n$. The self-consistent equation of the order parameter is
\begin{equation}
\Delta{\mathbf{r}}=V\sum_{n}u_{\mathbf{r}}^n v_{\mathbf{r}}^{n*} \tanh \left(\frac{E_n}{2k_BT}\right).
\end{equation}
We solve the set of BdG equations self-consitently via exact diagonalization method in real space. The system size of $24\times 24$ is used in the calculation and the convergence criterion of $\Delta_{\mathbf{r}}$ is set to be $10^{-4}$ in unit of nearest-neighbor hopping $t$. We find that the order parameter is uniform ( $\Delta_{\mathbf{r}}=\Delta$, where $\Delta$ is real) in the vicinity of the $\pi$-flux regime.  Our calculations (see Fig. \ref{lattice}b) show that as the on-site interaction $V$ increases from zero, the order parameter turn out to be non-zero at $V_c\approx 5.88t$ and the system undergoes a quantum phase transition from a semimetallic phase to a band insulator.

\begin{figure}[tbp]
\setlength{\unitlength}{2mm}
\begin{center}
\begin{picture}(20,28)
\put(-9,-5){\includegraphics[width=8cm,height=7.0cm]{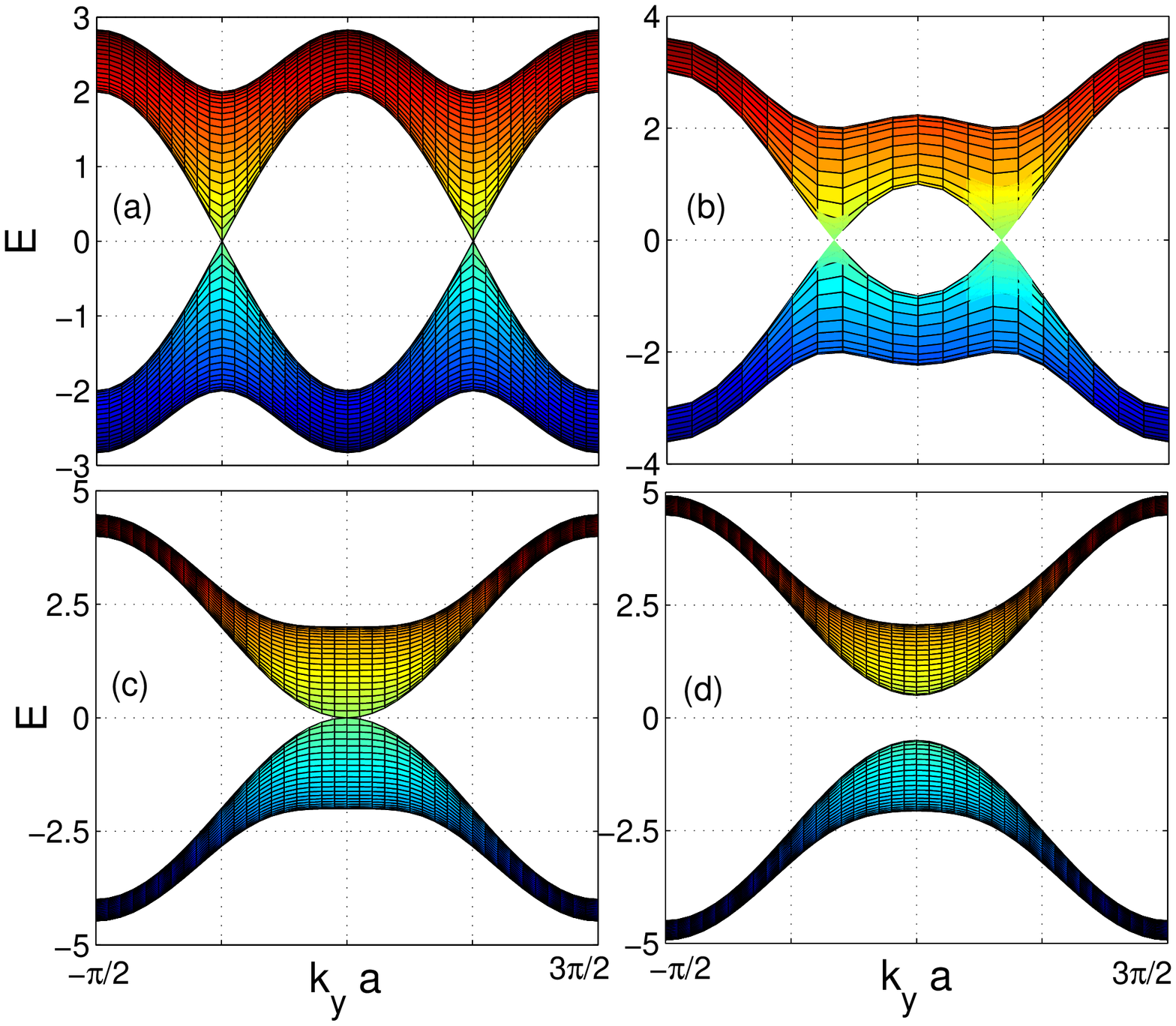}}
\end{picture}
\end{center}
\begin{center}
\begin{picture}(20,24)
\put(-9,-3){\scalebox{0.7}[0.7]{\includegraphics*[15,210][585,400]{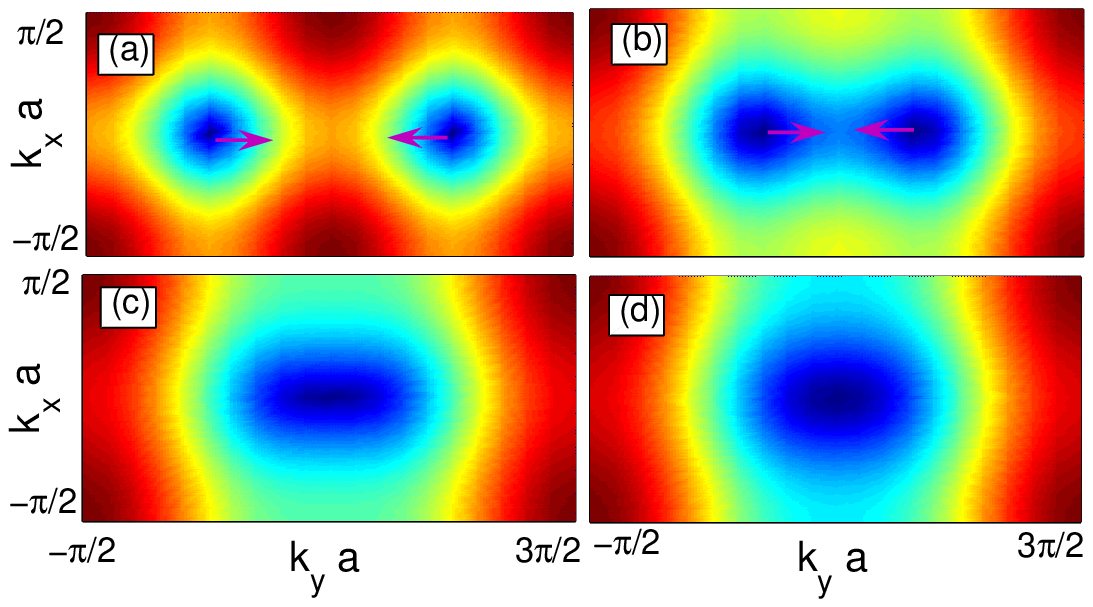}}}
\end{picture}
\end{center}
%\begin{center}
%\begin{picture}(20,17)
%\put(-9.5,-5){\resizebox{8cm}{!}{\includegraphics*[15,260][585,580]{densitplotb1.eps}}}
%\put(-6.8,-5){\resizebox{12cm}{!}{\includegraphics*[15,210][585,400]{DeltaU.eps}}}
%\end{picture}
%\end{center}
\caption{(Color online) Merging of Dirac points in the band structure of the system as $\Delta$ grows stronger($\Phi_{\alpha}=\Phi_{\beta}=\pi/2$). Top: Portrait of the energy spectrum in $k_x$ direction. Bottom: Variations process of the two Dirac points. (a) $\Delta=0$. There are two normal Dirac points. (b) $\Delta=1.0$. The two Dirac points moves closer. (c) $\Delta=2.0$. Two Dirac points merge into a single hybrid point, which signals the quantum phase transition. (d) $\Delta=2.5$. A gap is opened. \label{merge}}
\end{figure}

By the above mentioned self-consistent mean-field theory, we transform the original Hamiltonian into a quadratic form Eq.\;(\ref{Hc}). Implementing appropriate Fourier transformations, Eq.\;(\ref{Hc}) can be easily diagnalized in momentum space. The corresponding energy spectra are shown in Fig. \ref{merge}. As the mean-field order parameter $\Delta$ grows stronger, the two originally separate Dirac points (Fig. \ref{merge}a) will first move closer(Fig. \ref{merge}b), then merge into a single hybrid point at $k_h=(\frac{\pi}{2a},0)$. Around this hybrid point $p=k-k_h$, the low-energy properties of the system are accurately described by a  Dirac Hamiltonian
\begin{equation}
H_{eff}=\sum_{p} \Psi_p^{\dagger} H_D \Psi_p,  \;\;\; H_D=2 \sigma_y p_x-\sigma_x p_y^2
\end{equation}
where $\Psi_p=(c_{p\uparrow},c_{p\downarrow})^T$ is the relativistic spinor.
From this Hamiltonian we can see that the energy spectrum is linear in $k_x$ direction but quadratic in $k_y$ direction(Fig. \ref{merge}c).
The mergence of the two Dirac points signals a quantum phase transition from semimetallic phase to a band insulator \cite{Leticia,Montambaux}. If the order parameter $\Delta$ grows even stronger, a gap will be opened(Fig. \ref{merge}d), which indicates a band insulator phase. The self-consistent mean-field theory calculations show that the order parameter $\Delta$ remains to be zero as long as the on-site interaction is smaller than $U_c=5.88t$. Once the strength of the on-site interaction is over $U_c$, $\Delta$ will turn out to be a non-zero number which is not smaller than $\Delta_c=2$. This can be easily seen in Fig. \ref{lattice}b. Therefore, the spectrum shown in Fig. \ref{merge}b will not be observed in reality. We give out this spectrum in Fig. \ref{merge} just for comparison.

\begin{figure}[tbp]
\setlength{\unitlength}{2mm}
\begin{center}
\begin{picture}(20,25)
\put(-10,-4){\includegraphics[width=8cm]{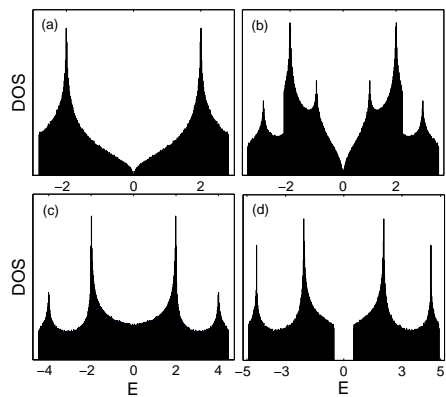}}
\end{picture}
\end{center}
\caption{Density of state (DOS) of the system vs. order parameter $\Delta$. (a) $\Delta=0$;(b) $\Delta=1.0$;(c) $\Delta=2.0$;(d) $\Delta=2.5$.  \label{dos}}
\end{figure}

\begin{figure}[tbp]
\setlength{\unitlength}{2mm}
\begin{center}
\begin{picture}(20,25)
\put(-13,-5){\includegraphics[width=9cm]{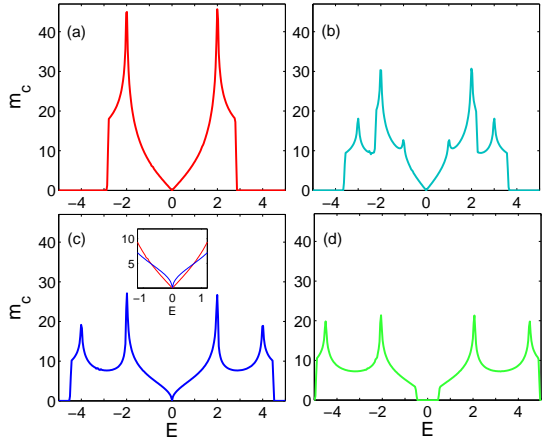}}
\end{picture}
\end{center}
\caption{(Color online) Cyclotron mass $m_c$ vs. $\Delta$. (a) $\Delta=0$;(b) $\Delta=1.0$;(c) $\Delta=2.0$;(d) $\Delta=2.5$. Inset: Comparison between (a) and (c) around the zero-energy point.   \label{mass}}
\end{figure}

To characterize the mergence of Dirac points more clearly, we calculate density of state(DOS) and cyclotron mass $m_c$ for different strength of on-site interaction (i.e. different values of $\Delta$ correspondingly). The results are shown in Fig. \ref{dos} and Fig. \ref{mass}.
% here I will give out the formula of the mass,and list out the corresponding references
For the density of states in Fig. \ref{dos}, we find that dramatic difference appeared at the occasion when two Dirac points merge into a hybrid one. While the DOS displays two peaks for the case of $\Delta=0$(Fig. \ref{dos}a), there are four peaks in the case of $\Delta=2$(Fig. \ref{dos}c). Moreover, for $\Delta=2$ (Fig. \ref{dos}c), DOS turns out to be non-zero at $E=0$. These obvious differences are supposed to be good indexes for the mergence of the Dirac points. In Fig. \ref{mass}, we give out the cyclotron mass $m_c$ for different values of $\Delta$, correspondingly, different strengthes of on-site interaction. Comparing Fig. \ref{mass}a and Fig. \ref{mass}c, we find that besides the difference in the numbers of peaks, the curve of $m_c$ changes from concave to convex around the zero-energy point as shown in the inset of Fig. \ref{mass}c. This change signals a quantum phase transition between the semimetalic phase and band insulator phase.

From Fig. \ref{lattice}b, we see that by increasing the strength of the on-site interaction, the mean-field order parameter can be tuned. And it's well-known that the on-site interaction can be controlled conveniently by Feshbach resonances. That is, the mean-field order parameter could be modified by Feshbach resonaces indirectly. Therefore, this exotic mergence of the Dirac points induced by on-site interaction is supposed to be observed in ultracold fermi gas experiments. This is quite different from the works \cite{Leticia,Montambaux}, where they engineer Dirac points by tuning the nearest-neighbor tunneling \cite{Montambaux} or by tuning the geometry of the optical lattices\cite{Leticia}. For the concrete realization of the experiment, one may use $^{40}K$ atoms in $F=9/2$ or $F=7/2$ hyperfine manifolds or $^6Li$ with $F=1/2$.

%\section{Conclusion}
%\emph{Conclusion}---
Dirac point plays a crucial role in many interesting phenomena in condensed-matter physics, for example, the massless electrons in graphene. In this article, we investigate a two-component(color) ultracold fermi gas which is loaded in a square optical lattice. The addition of non-Abelian artificial gauge filed gives rise of Dirac points in the energy spectrum of the system. We study the stability of Dirac points against the repulsive on-site interaction. Our calculations show that the Dirac points can be very stable under on-site interaction smaller than $U_c=5.88t$.  At $U_c$, the Dirac points turn to be non-stable and merge into a hybrid point. The final hybrid point is linear in one direction but quadratic in the perpendicular direction. This mergence of the Dirac points denotes a quantum phase transition from semimetallic phase to a band insulator. This exotic phenomena is supposed to observed in ultracold fermi gas experiments nowadays.

This work is supported by the NFRP (2011CB921503) and the NNSF of China (Grants No. 11075020 and No. 91021021). L. Wang appreciate very much the help of N.N. Hao for valuable discussions on code writing.

\end{document}